\documentclass[prd,referee]{revtex4-1}
\usepackage{eurosym}
\usepackage{graphicx}
\usepackage{amsmath}
\usepackage{bm}
\usepackage{color}
\usepackage{dcolumn}%
\usepackage{bm}%

\def\be{\begin{equation}}
\def\ee{\end{equation}}
\def\bea{\begin{eqnarray}}
\def\eea{\end{eqnarray}}

\def\bee{\beta}
\def\si{\sigma}

\begin{document}

\title{Jacobi stability analysis of the Lorenz system}
\author{Tiberiu Harko}
\email{t.harko@ucl.ac.uk} \affiliation{Department of
Mathematics, University College London, Gower Street, London WC1E
6BT, United Kingdom}
\author{Chor Yin Ho}
\email{chor.yin.ho@polyu.edu.hk} \affiliation{Department of
Applied Mathematics, Polytechnic University, Hong Kong}
\author{Chun Sing Leung}
\email{chun-sing-hkpu.leung@polyu.edu.hk} \affiliation{Department of
Applied Mathematics, Polytechnic University, Hong Kong}
\author{Stan Yip}
\email{stan.yip@polyu.edu.hk} \affiliation{Department of
Applied Mathematics, Polytechnic University, Hong Kong}
\begin{abstract}
We perform the study of the stability of the Lorenz system by using
the Jacobi stability analysis, or the Kosambi-Cartan-Chern (KCC)
theory. The Lorenz model plays an important role for understanding
hydrodynamic instabilities and the nature of the turbulence, also
representing a non-trivial testing object for studying non-linear
effects. The KCC theory represents a powerful mathematical method
for the analysis of dynamical systems. In this approach we describe
the evolution of the Lorenz system in geometric terms, by
considering it as a geodesic in a Finsler space. By associating a
non-linear connection and a Berwald type connection, five
geometrical invariants are obtained, with the second invariant
giving the Jacobi stability of the system. The Jacobi (in)stability
is a natural generalization of the (in)stability of the geodesic
flow on a differentiable manifold endowed with a metric (Riemannian
or Finslerian) to the non-metric setting. In order to apply the KCC
theory we reformulate the Lorenz system as a set of two second order
non-linear differential equations. The geometric invariants
associated to this system (nonlinear and Berwald connections), and
the deviation curvature tensor, as well as its eigenvalues,  are
explicitly obtained. The Jacobi stability of the equilibrium points
of the Lorenz system is studied, and the condition of the stability
of the equilibrium points is obtained. Finally, we consider the time
evolution of the components of the deviation vector near the
equilibrium points.
\end{abstract}


\date{\today}

\maketitle

\section{Introduction}

Continuously time evolving dynamical systems are one of the basic theoretical tools for
modeling the evolution of natural phenomena in every branch of physics, chemistry, or biology. Their
usefulness in scientific/engineering applications is determined by their predictive power,  which, in turn,
strongly depends on the stability of their solutions. Since in the measured initial conditions in a physical system some uncertainty
inevitably does exist, a physically meaningful
mathematical model must offer an understanding of the possible evolution
of the deviations of the trajectories of the studied dynamical system from a given reference trajectory. Note that a local understanding
of the stability is as important as the global evolution and control
of late-time deviations. From a mathematical point of view the global stability of the solutions of the dynamical systems is described by the well studied theory of Lyapounov stability. In this approach the fundamental quantities are the Lyapunov exponents, measuring exponential deviations from the given trajectory \cite{1,2}. It is usually very difficult to analytically determine the  Lyapounov exponents, and therefore various numerical methods  for their calculation have been proposed, and are used in various situations \cite{3}-\cite{11}.  On the other hand the local stability of solutions of dynamical systems is much less understood.

Even that the methods of the Lyapounov stability analysis are well established, it would be interesting to study the stability of the  dynamical system
from different points of view, and to compare the results with the corresponding
Lyapunov exponents analysis. Such an alternative approach to the study of the dynamical systems is represented by the so-called geometrodynamical approach, which was initiated in the pioneering work of Kosambi \cite{Ko33}, Cartan \cite{Ca33} and Chern \cite{Ch39}. The Kosambi-Cartan-Chern (KCC) approach is inspired by the geometry of the Finsler spaces. Its basic idea is to consider that there is a one to one correspondence between a second order dynamical system  and the geodesic equations in an associated Finsler space (for a recent review of the KCC theory see \cite{rev}). The KCC theory is a differential
geometric theory of the variational equations for the deviations of the whole trajectory to nearby ones \cite{An00}. In this geometrical description of the dynamical systems one associates a non-linear connection, and a Berwald type connection to the differential system, and five geometrical
invariants are obtained. The second invariant, also called the curvature deviation tensor,   gives the Jacobi stability of
the system \cite{rev, An00, Sa05,Sa05a}. The KCC theory has been applied for the study
of different physical, biochemical or technical systems (see \cite{Sa05,
Sa05a, An93, YaNa07,Ha1,Ha2}).

An alternative geometrization method for dynamical systems was proposed in \cite{Pet1} and \cite{Kau}, and further investigated in \cite{Pet0}-\cite{Pet4}. Specific applications for the Henon-Heiles system and Bianchi type IX cosmological models were also considered. In particular, in \cite{Pet0} a theoretical framework devoted to a geometrical description of the behavior of
dynamical systems and their chaotic properties was developed.

In the Riemmannian geometric approach to dynamical systems one starts with the well-known results that the flow associated with a time dependent Hamiltonian $H=\delta ^{ab}p_ap_b/2+V\left(x^a\right)$ can be reformulated as a geodesic flow in a curved, but conformally flat,
manifold \cite{Kau}. By introducing a metric of the form $ds^2=W\left(x^a\right)\delta _{ab}dx^adx^b$, with the conformal factor given by $W\left(x^a\right)=E-V\left(x^a\right)$, where $E$ is  the conserved energy associated with the time-independent $H$, it follows that the geodesic equation for motion in the metric $g_{ab} = W\left(x^a\right)\delta _{ab}$ is completely equivalent to the
Hamilton equations $dx^a/dt=\partial H/\partial p_a$, $dp_a/dt=-\partial H/\partial x_a$ \cite{Kau}. This implies that the confluence or divergence of nearby trajectories $x^a(s)$ and $[x+\xi]^a(s)$ is determined by the Jacobi equation, i.e., the equation of geodesic deviation, which takes
the form
\be
\frac{D^2\xi ^a}{Ds^2} = −R^a_{bcd}u^bu^d\xi ^c \equiv - K^a_c \xi ^c,
\ee
where $R^a_{bcd}$ is the Riemann tensor associated with $g_{ab}$, and $D/Ds = u^a\nabla _a$ denotes a directional
derivative along $u^a = dx^a/ds$. Linear stability for the trajectory $x^a(s)$ is thus related to $R^a_{bcd}$ or, more exactly, to the curvature $K^a_c$. If, for example, $R^a_{bcd}$ is everywhere
negative, so that $K^a_c$ always has one or more negative eigenvalues, the trajectory must be
linearly unstable \cite{Kau}.

 The global and local stability of solutions of continuously evolving dynamical systems was reconsidered,
from a geometric perspective, in \cite{Punzi}. It was shown that an unambiguous definition of stability generally requires the
choice of additional geometric structures that are not intrinsic to the dynamical system itself.

One of the most studied non-linear differential equations system is the Lorenz system \cite{Lorenz, Lorenz1, Lorenz2}, which has  been proposed as a possible description of the mechanism of the transition to weak turbulence  in natural convection, by simulating thermal convection in the atmosphere, and modeling turbulent convective flow. An alternate view for the emergence of chaos in Lorenz-like systems was considered in \cite{PRE1}.    The effects of noise on the Lorenz equations in the parameter regime admitting two stable fixed point solutions and a strange attractor was studied in \cite{PRL1}.
Absolute periods and symbolic names were assigned to stable and unstable periodic orbits of the Lorenz system in \cite{PRE2}.   Whether unstable periodic orbits  associated with a strange attractor may predict the occurrence of the robust sharp peaks in histograms of some experimental chaotic time series for the Lorenz equations was investigated in \cite{PRL2}.
 A nonlinear feedback approach for controlling the Lorenz equation was proposed in \cite{PRE3}.  A particular return map for a class of low dimensional chaotic models called Kolmogorov - Lorenz systems where studied from the viewpoint of energy and Casimir balance in \cite{Pel}, by using a general Hamiltonian description.

The possibility that the Lorenz system can be discussed in terms of the KCC-theory in Finsler space was pointed out in \cite{T0}, but without presenting any concrete analysis.
A geometric viewpoint of the Lorenz system, based on a theory of tangent bundle, was proposed in \cite{T1}. By
introducing the geometrical viewpoints of second order system governed by Euler - Poincar\'{e} equation or
Lie - Poisson equation, the geometrical invariants of the Lorenz system have been obtained. It was shown that a torsion
tensor, as one of geometrical invariants, relates to the chaotic behavior, characterized by the Rayleigh
number, and results from the decomposition from the second order system to the tangent space (state
space) and base space (configuration space), respectively.

It is the purpose of the present paper to consider a full analysis of the Lorenz equations in the framework of the KCC theory. As a first step in this approach, the Lorenz system is reformulated as a set of two second order non-linear differential equations. The geometric invariants associated to this system (nonlinear and Berwald connections), and the components of the deviation curvature tensor are explicitly obtained, as well as its eigenvalues. The Jacobi stability of the equilibrium points of the Lorenz system is studied, and the condition of the stability of the equilibrium points is obtained. Finally, we consider the time evolution of the components of the deviation vector near the equilibrium points.

The present paper is organized as follows. In Section \ref{sect3} we reformulate the Lorenz system as a set of two second order differential equations. The basics of the KCC theory to be used in the sequel are presented in Section~\ref{kcc}. The Jacobi stability of the Lorenz system is considered in Section~\ref{sect4}. The time evolution of the components of the deviation vector is studied in Section~\ref{sect5}. We discuss and conclude our results in Section~\ref{sect6}.

\section{The Lorenz equations}\label{sect3}

In order to discuss the application of the KCC theory to systems of
differential equations connected to fluid mechanics, we consider the Lorenz
system of three nonlinear ordinary differential equations, given by \cite%
{Lorenz}
\begin{equation}
\frac{1}{\sigma }\frac{dX}{dt}=-X+Y,  \label{L1}
\end{equation}%
\begin{equation}
\frac{dY}{dt}=-XZ+\rho X-Y,  \label{L2}
\end{equation}%
\begin{equation}
\frac{dZ}{dt}=XY-\beta Z,  \label{L3}
\end{equation}%
where $\sigma $, $\rho $ and $\beta $ are some free parameters. From a mathematical point of view these
ordinary differential equations represent an approximation to a system of partial
differential equations describing finite amplitude convection in a fluid
layer heated from below. The
Lorenz system results if the unknown functions in the partial
differential equations are expanded in Fourier series, and all the
resulting Fourier coefficients are set equal to zero except three \cite{Lorenz}. The parameters $\sigma $, $\rho $, and $\beta $
can be interpreted from a physical point of view as  the Prandl number, the Rayleigh number (suitably
normalized), and the wave length number, respectively. Although the Lorenz set of equations is deterministic, its solution shows chaotic behavior for $\rho > \rho _{crit}=\sigma (\sigma +\beta +3)/(\sigma -\beta -1)$, and $\sigma >\beta +1$ \cite{Lorenz, Rossler}.

\subsection{ Second order differential equations
formulations of the Lorenz system}

From Eq. (\ref{L1}) we can express $Y$ as
\begin{equation}
Y=X+\frac{1}{\sigma }\dot{X}.  \label{Y}
\end{equation}

By substituting $Y$ into Eqs. (\ref{L2}) we obtain
\begin{equation}
\ddot{X}+(1+\sigma )\dot{X}+\sigma XZ+\sigma \left( 1-\rho \right) X=0.
\label{F1}
\end{equation}

By taking the derivative of Eq. (\ref{L3}) with respect to the time  we find
\bea
\ddot{Z}&=&\dot{X}Y+X\dot{Y}-\beta \dot{Z}=\dot{X}\left( X+\frac{1}{\sigma }%
\dot{X}\right) +X\left( \dot{X}+\frac{1}{\sigma }\ddot{X}\right) -
\beta %
\left[ X\left( X+\frac{1}{\sigma }\dot{X}\right) -\beta Z\right] .
\label{F2i}
\eea

By substituting $\ddot{X}$ as given by Eq. (\ref{F1}) we obtain the
following equation for $\ddot{Z}$,
\bea
&&\ddot{Z}+\left( \frac{1+\sigma }{\sigma }+\frac{\beta }{\sigma }-2\right) X%
\dot{X}-\frac{1}{\sigma }\dot{X}^{2}+\left( 1-\rho +\beta \right)
X^{2}+
X^{2}Z-\beta ^{2}Z=0.  \label{F2}
\eea

At this moment we change the notation so that
\be
X=X^{1}, \dot{X}=Y^{1}, %
Z=X^{2}, \dot{Z}=Y^{2},
\ee
and
\be
Y=X^{3},
\ee
respectively. Hence the Lorenz system is equivalent with the following system of second order equations,
\begin{equation}
\frac{d^{2}X^{1}}{dt^{2}}+\left( 1+\sigma \right) Y^{1}+\sigma
X^{1}X^{2}+\sigma \left( 1-\rho \right) X^{1}=0,  \label{Ff1}
\end{equation}
\begin{eqnarray}\label{Ff2}
&&\frac{d^{2}X^{2}}{dt^{2}}+\left( \frac{1+\sigma }{\sigma }+\frac{\beta }{%
\sigma }-2\right) X^{1}Y^{1}-\frac{1}{\sigma }\left( Y^{1}\right)
^{2}+\left( 1-\rho +\beta \right) \left( X^{1}\right) ^{2}+\left(
X^{1}\right) ^{2}X^{2}-\beta ^{2}X^{2}=0. \nonumber\\
\end{eqnarray}

Once $X^{1}$ and $X^{2}$ are known, the variable $X^{3}$ can be obtained as
\begin{equation}
X^{3}=X^{1}+\frac{1}{\sigma }Y^{1}.  \label{Y3}
\end{equation}

\section{Kosambi-Cartan-Chern (KCC) theory and Jacobi stability}\label{kcc}

In the present Section we summarize the basic concepts and results
of the KCC theory (for a detailed presentation see \cite{rev} and
\cite{An00}).

Let $\mathcal{M}$ be a real, smooth $n$-dimensional manifold and let $T%
\mathcal{M}$ be its tangent bundle. On an open connected subset $\Omega $ of the Euclidian $(2n+1)$
dimensional space $R^{n}\times R^{n}\times R^{1}$  we introduce a $2n+1$ dimensional coordinates
system $\left(x^i,y^i,t\right)$, $i=1,2,...,n$, where $\left( x^{i}\right) =\left(
x^{1},x^{2},...,x^{n}\right) $, $\left( y^{i}\right) =\left(
y^{1},y^{2},...,y^{n}\right) $ and $t$ is the usual time coordinate. The coordinates $y^i$ are defined as
\begin{equation}
y^{i}=\left( \frac{dx^{1}}{dt},\frac{dx^{2}}{dt},...,\frac{dx^{n}}{dt}%
\right) .
\end{equation}

In the following we assume that the time coordinate $t$ is an absolute invariant. Therefore the only
admissible coordinate transformations are
\begin{equation}
\tilde{t}=t,\tilde{x}^{i}=\tilde{x}^{i}\left( x^{1},x^{2},...,x^{n}\right)
,i\in \left\{1 ,2,...,n\right\} .  \label{ct}
\end{equation}

Following \cite{Punzi}, we define a deterministic dynamical systems  as a set of formal rules describing the evolution of
points in a set $S$ with respect to an external, discrete, or
continuous time parameter $t\in T$. More
precisely, a dynamical system is a map \cite{Punzi}
\be
\phi:T \times S \rightarrow S, (t,x)\mapsto \phi (t,x),
\ee
which satisfies the condition $\phi (t , \cdot) \circ \phi (s , \cdot)=\phi (t+s , \cdot)$, $\forall t ,s\in  T$.
For realistic dynamical systems additional structures must be added to this definition.

In many situations the equations of motion of a dynamical system can be derived from a
Lagrangian $L$ via the Euler-Lagrange equations,
\begin{equation}
\frac{d}{dt}\frac{\partial L}{\partial y^{i}}-\frac{\partial L}{\partial
x^{i}}=F_{i},i=1,2,...,n,  \label{EL}
\end{equation}%
where $F_{i}$, $i=1,2,...,n$, is the external force. The
triplet $\left( M,L,F_{i}\right) $ is called a Finslerian mechanical system
 \cite{MiFr05,MHSS}. For a regular Lagrangian $L$, the Euler-Lagrange equations
introduced in Eq.~(\ref{EL}) are equivalent to a system of second-order ordinary (usually nonlinear)
differential equations
\begin{equation}
\frac{d^{2}x^{i}}{dt^{2}}+2G^{i}\left( x^{j},y^{j},t\right) =0,i\in \left\{
1,2,...,n\right\} ,  \label{EM}
\end{equation}%
where each function $G^{i}\left( x^{j},y^{j},t\right) $ is $C^{\infty }$ in
a neighborhood of some initial conditions $\left( \left( x\right)
_{0},\left( y\right) _{0},t_{0}\right) $ in $\Omega $.


The basic idea of the KCC theory is to start from an arbitrary system of second-order
differential equations of the form (\ref{EM}), with no \textit{a priori}
given Lagrangean function assumed, and study the behavior of its
trajectories by analogy with the trajectories of the Euler-Lagrange
system.

To analyze the geometry associate to the dynamical system defined by Eqs.~(\ref{EM}),
as a first step we introduce a nonlinear connection $N$ on $M$, with
coefficients $N_{j}^{i}$, defined as \cite{MHSS}
\begin{equation}
N_{j}^{i}=\frac{\partial G^{i}}{\partial y^{j}}.
\end{equation}

The nonlinear connection can be understood geometrically in terms of a
dynamical covariant derivative $\nabla ^N$ \cite{Punzi}: for two vector fields $v$, $w$
defined over a manifold $M$, we introduce the covariant derivative $\nabla ^N$ as
\be\label{con}
\nabla _v^Nw=\left[v^j\frac{\partial }{\partial x^j}w^i+N^i_j(x,y)w^j\right]\frac{\partial }{\partial x^i}.
\ee

For $N_i^j(x,y)=\Gamma _{il}^j(x)y^l$, Eq.~(\ref{con}) reduces to the definition of the covariant derivative for the special case of a standard linear connection, as defined in Riemmannian geometry.

For the non-singular coordinate transformations introduced through Eqs.~(\ref{ct}), we
define the KCC-covariant differential of a vector field $\xi ^{i}(x)$ on the
open subset $\Omega \subseteq R^{n}\times R^{n}\times R^{1}$ as \cite%
{An93,An00,Sa05,Sa05a}
\begin{equation}
\frac{D\xi ^{i}}{dt}=\frac{d\xi ^{i}}{dt}+N_{j}^{i}\xi ^{j}.  \label{KCC}
\end{equation}

For $\xi ^{i}=y^{i}$ we obtain
\begin{equation}
\frac{Dy^{i}}{dt}=N_{j}^{i}y^{j}-2G^{i}=-\epsilon ^{i}.
\end{equation}
The contravariant vector field $\epsilon ^{i}$ on $\Omega $ is called
the first KCC invariant.

We vary now the trajectories $x^{i}(t)$ of the system (\ref{EM}) into
nearby ones according to
\begin{equation}
\tilde{x}^{i}\left( t\right) =x^{i}(t)+\eta \xi ^{i}(t),  \label{var}
\end{equation}
where $\left| \eta \right| $ is a small parameter, and $\xi ^{i}(t)$ are the
components of a contravariant vector field defined along the path $%
x^{i}(t)$. Substituting Eqs. (\ref{var}) into Eqs. (\ref{EM}) and taking the
limit $\eta \rightarrow 0$ we obtain the deviation equations in the form \cite%
{An93,An00,Sa05,Sa05a}
\begin{equation}
\frac{d^{2}\xi ^{i}}{dt^{2}}+2N_{j}^{i}\frac{d\xi ^{j}}{dt}+2\frac{\partial
G^{i}}{\partial x^{j}}\xi ^{j}=0.  \label{def}
\end{equation}

Eq.~(\ref{def}) can be reformulate in the
covariant form with the use of the KCC-covariant differential as
\begin{equation}
\frac{D^{2}\xi ^{i}}{dt^{2}}=P_{j}^{i}\xi ^{j},  \label{JE}
\end{equation}
where we have denoted
\begin{equation}
P_{j}^{i}=-2\frac{\partial G^{i}}{\partial x^{j}}-2G^{l}G_{jl}^{i}+ y^{l}%
\frac{\partial N_{j}^{i}}{\partial x^{l}}+N_{l}^{i}N_{j}^{l}+\frac{\partial
N_{j}^{i}}{\partial t},
\end{equation}
and we have introduced the Berwald connection $G_{jl}^{i}$, defined as \cite{rev, An00, An93,MHSS,Sa05,Sa05a}
\be
G_{jl}^{i}\equiv \frac{\partial N_{j}^{i}}{\partial y^{l}}.
\ee

 $P_{j}^{i}$ is called the second KCC-invariant or the
deviation curvature tensor, while Eq.~(\ref{JE}) is called the
Jacobi equation. When the system (\ref{EM}) describes the
geodesic equations, Eq.~(\ref{JE}) is the Jacobi field equation, in either Riemann or Finsler geometry.   .

The trace $P$ of the curvature deviation tensor is obtained as
\be
P=P_{i}^{i}=-2\frac{\partial G^{i}}{\partial x^{i}}-2G^{l}G_{il}^{i}+ y^{l}%
\frac{\partial N_{i}^{i}}{\partial x^{l}}+N_{l}^{i}N_{i}^{l}+\frac{\partial
N_{i}^{i}}{\partial t}.
\ee

The third, fourth and fifth invariants of the system (\ref{EM}) are defined as
\cite{An00}
\begin{equation}\label{31}
P_{jk}^{i}\equiv \frac{1}{3}\left( \frac{\partial P_{j}^{i}}{\partial y^{k}}-%
\frac{\partial P_{k}^{i}}{\partial y^{j}}\right) ,P_{jkl}^{i}\equiv \frac{%
\partial P_{jk}^{i}}{\partial y^{l}},D_{jkl}^{i}\equiv \frac{\partial
G_{jk}^{i}}{\partial y^{l}}.
\end{equation}

The third invariant $P_{jk}^{i}$ can be interpreted geometrically as a torsion tensor. The fourth and
fifth invariants $P_{jkl}^{i}$ and $D_{jkl}^{i}$ are called the Riemann-Christoffel curvature tensor, and the
Douglas tensor, respectively \cite{rev, An00}. In a Berwald space these tensors
always exist. In the KCC theory they describe the geometrical properties and interpretation of a system of
second-order differential equations.

Alternatively, we can introduce another definition for the third and fourth KCC invariants, as \cite{T0}
\be\label{32}
B_{jk}^i=\frac{\delta N_j^i}{\delta x^k}-\frac{\delta N_k^i}{\delta x^j},
\ee
where
\be
\frac{\delta }{\delta x^i}=\frac{\partial}{\partial x^i}-N_i^j\frac{\partial }{\partial y^j}.
\ee

The fourth KCC invariant can then be defined as
\be
B_{jkl}^i=\frac{\partial B_{kl}^i}{\partial y^j}.
\ee

In many physical, chemical or biological applications we are interested in the behavior of the
trajectories of the dynamical system (\ref{EM}) in a vicinity of a point $x^{i}\left(
t_{0}\right) $. For simplicity in the following we take $t_{0}=0$. We
consider the trajectories $x^{i}=x^{i}(t)$ as curves in the Euclidean space $%
\left( R^{n},\left\langle .,.\right\rangle \right) $, where $\left\langle
.,.\right\rangle $ is the canonical inner product of $R^{n}$. For the
deviation vector $\xi $ we assume that it obeys  the initial conditions $%
\xi \left( 0\right) =O$ and $\dot{\xi}\left( 0\right) =W\neq O$, where $O\in
R^{n}$ is the null vector \cite{rev, An00, Sa05,Sa05a}.


Thus, we introduce the following description of the focusing tendency of the trajectories around $t_{0}=0$: if $\left| \left| \xi \left( t\right) \right| \right|
<t^{2}$, $t\approx 0^{+}$, the trajectories are bunching together. On the other hand, if $%
\left| \left| \xi \left( t\right) \right| \right| >t^{2}$, $t\approx 0^{+}$,
the trajectories are dispersing \cite{rev, An00, Sa05,Sa05a}. The focusing tendency of the trajectories can be also characterized in terms of the deviation
curvature tensor in the following way: The trajectories of the system of equations (\ref{EM}) are
bunching together for $t\approx 0^{+}$ if and only if the real part of the
eigenvalues of the deviation tensor $P_{j}^{i}\left( 0\right) $ are strictly negative. The trajectories
are dispersing if and only if the real part of the eigenvalues of $%
P_{j}^{i}\left( 0\right) $ are strictly positive \cite{rev, An00, Sa05,Sa05a}.

Based on the above considerations we  define the concept of the Jacobi stability for a
dynamical system as follows \cite{rev, An00,Sa05,Sa05a}:

\textbf{Definition:} If the system of differential equations Eqs.~(\ref{EM})
satisfies the initial conditions $\left| \left| x^{i}\left( t_{0}\right) -%
\tilde{x}^{i}\left( t_{0}\right) \right| \right| =0$, $\left| \left| \dot{x}%
^{i}\left( t_{0}\right) -\tilde{x}^{i}\left( t_{0}\right) \right| \right|
\neq 0$, with respect to the norm $\left| \left| .\right| \right| $ induced
by a positive definite inner product, then the trajectories of Eqs.~(\ref{EM})
are Jacobi stable if and only if the real parts of the eigenvalues of the
deviation tensor $P_{j}^{i}$ are strictly negative everywhere. Otherwise, the trajectories are Jacobi
unstable.

Graphically, the focussing behavior of the trajectories near the origin is shown in
Fig.~\ref{pict1}.
\begin{widetext}
\begin{centering}
\begin{figure*}[htp]
\centering
\includegraphics[height=6cm,width=12cm]{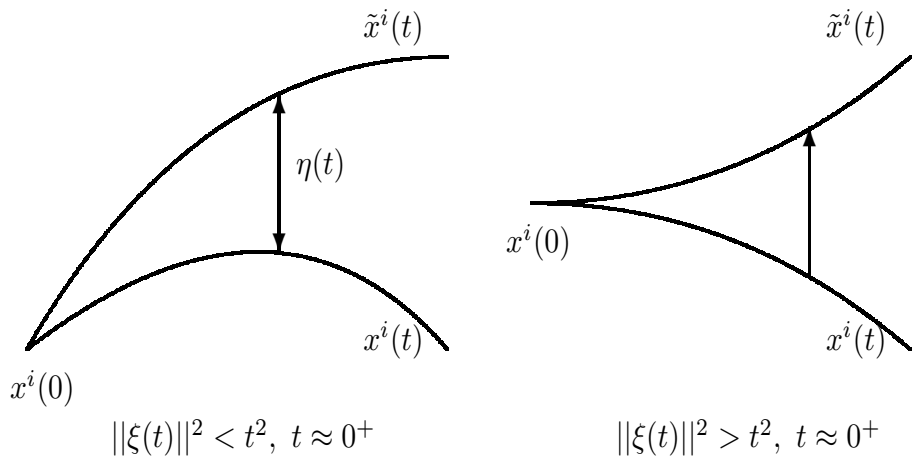} 
\caption{Behavior of the trajectories near zero.}
\label{pict1}
\end{figure*}
\end{centering}
\end{widetext}

The curvature deviation tensor can be written in a matrix form as
\begin{equation}
P_{j}^{i}=\left(
\begin{array}{c}
P_{1}^{1}\;\;\;\;P_{2}^{1} \\
P_{1}^{2}\;\;\;P_{2}^{2}%
\end{array}%
\right) ,
\end{equation}%
with the eigenvalues given by
\begin{equation}
\lambda _{\pm }=\frac{1}{2}\left[ P_{1}^{1}+P_{2}^{2}\pm \sqrt{\left(
P_{1}^{1}-P_2^2\right) ^{2}+4P_{2}^{1}P_{1}^{2}}\right] .
\end{equation}

The eigenvalues of the curvature deviation tensor are the solutions of the quadratic equation
\be\label{eql}
\lambda ^2-\left(P_1^1+P_2^2\right)\lambda +\left(P_1^1P_2^2-P_2^1P_1^2\right)=0.
\ee

In order to obtain  the signs of the eigenvalues of the curvature deviation tensor we use the Routh-Hurwitz criteria \cite{RH}, according to which all of the roots of the polynomial $P(\lambda)$ are negatives or have negative real parts if the determinant of all Hurwitz
matrices ${\rm det}\;H_j$, $j=1,2,..,n$, are positive. For $n=2$, corresponding to the case of Eq.~(\ref{eql}), the Routh-Hurwitz criteria simplify to
\be
P_1^1+P_2^2<0,\;\;P_1^1P_2^2-P_2^1P_1^2>0.
\ee

$\lambda _{\pm}$ describe the curvature properties along a given
geodesic. Hence we can characterize the way the geodesic explore the Finsler manifold through the (half) of the Ricci curvature scalar along the flow, $\kappa $, and the anisotropy $\theta $, defined as \cite{PR}
\be
\kappa =\frac{1}{2}\left(\lambda _{+}+\lambda _{-}\right)=\frac{P}{2}=\frac{P_1^1+P_2^2}{2},
\ee
and
\be
\theta =\frac{1}{2}\left(\lambda _{+}-\lambda _{-}\right)=\frac{\sqrt{\left(
P_{1}^{1}-P_2^2\right) ^{2}+4P_{2}^{1}P_{1}^{2} }}{2},
\ee
respectively.

\section{Jacobi stability of the Lorenz system}\label{sect4}

In the present Section we use the KCC approach for the study of the dynamical properties of the Lorenz system. We explicitly obtain  the non-linear and Berwald connections, and the deviation curvature tensors for the Lorenz system. The eigenvalues of the deviation curvature tensor are also obtained, and we study their properties in the equilibrium points of the Lorenz system. The study of the sign of the eigenvalues allows us to formulate a basic theorem giving the Jacobi stability properties of the fixed points of the Lorenz system.

\subsection{The non-linear and Berwald connections, and the KCC invariants
of the Lorenz system}

The Lorenz system can be formulated as a second order differential system, given by two equations of the form
\be
\frac{d^2X^i}{dt^2}+2G^i\left(X^i,Y^i\right)=0, i=1,2.
\ee

From Eqs. (\ref{Ff1}) and (\ref{Ff2}) it follows immediately that
\begin{equation}
G^{1}\left( X^{1},X^{2},Y^{1}\right) =\frac{1}{2}\left[ \left( 1+\sigma
\right) Y^{1}+\sigma X^{1}X^{2}+\sigma \left( 1-\rho \right) X^{1}\right] ,
\label{G1}
\end{equation}%
and
\bea
G^{2}\left( X^{1},X^{2},Y^{1}\right) &=&\frac{1}{2}\Bigg[ \left( \frac{%
1+\sigma }{\sigma }+\frac{\beta }{\sigma }-2\right) X^{1}Y^{1}-
\frac{1}{\sigma }\left( Y^{1}\right) ^{2}+\left( 1-\rho +\beta \right) \left(
X^{1}\right) ^{2}+\nonumber\\
&&\left( X^{1}\right) ^{2}X^{2}-\beta ^{2}X^{2}\Bigg] ,
\label{G2}
\eea
respectively. Therefore we first obtain the components of the non-linear
connection as
\begin{eqnarray}
N_{1}^{1} &=&\frac{\partial G^{1}\left( X^{1},X^{2},Y^{1}\right) }{\partial
Y^{1}}=\frac{1}{2}\left( 1+\sigma \right) ,N_{2}^{1}=0,  \label{N} \\
N_{1}^{2} &=&\frac{\partial G^{2}\left( X^{1},X^{2},Y^{1}\right) }{\partial
Y^{1}}=\frac{1}{2}\left( \frac{1+\sigma }{\sigma }+\frac{\beta }{\sigma }%
-2\right) X^{1}-\frac{1}{\sigma }Y^{1},N_{2}^{2}=0.
\end{eqnarray}

For the components of the Berwald connection we obtain
\begin{equation}
G_{11}^{1}=G_{12}^{1}=G_{21}^{1}=G_{22}^{1}=0,
G_{11}^{2}=-\frac{1}{\sigma },G_{12}^{2}=G_{21}^{2}=G_{22}^{2}=0.
\end{equation}

The components of the first KCC invariant of the Lorenz system are given by
\be
\epsilon ^1=\frac{1+\sigma}{2}Y^1+\sigma X^1X^2+\sigma (1-\rho )X^1,
\ee
and
\bea
\epsilon ^2=\frac{1}{2}\left(\frac{1+\beta}{\sigma }-1\right)X^1Y^1+\left(1-\rho +\beta\right)\left(X^1\right)^2+
\left(X^1\right)^2X^2-\beta ^2X^2,
\eea
respectively.

The components of the curvature deviation tensor of the Lorenz system are
given by
\begin{equation}
P_{1}^{1}=-\sigma X^{2}-\sigma \left( 1-\rho \right) +\frac{1}{4}\left(
1+\sigma \right) ^{2},
\end{equation}
\begin{equation}
P_{2}^{1}=-\sigma X^{1},
\end{equation}
\bea
P_{1}^{2}&=&\left(1-\frac{\beta }{2 \sigma }\right)Y^1 +
\left[\frac{1-\sigma ^2-7 \beta  \sigma +\beta +4 (\rho -1) \sigma }{4 \sigma }\right]X^1 -
X^1X^2,
\eea
\begin{equation}
P_{2}^{2}=-\left( X^{1}\right) ^{2}+\beta ^{2}.
\end{equation}
For the trace of the curvature deviation tensor we obtain
\be
P=P_1^1+P_2^2=-\left( X^{1}\right) ^{2}-\sigma X^{2}-\sigma \left( 1-\rho \right) +\frac{1}{4}\left(
1+\sigma \right) ^{2}+\beta ^{2}.
\ee

\begin{figure*}[htp]
\begin{centering}
\includegraphics[width=8cm]{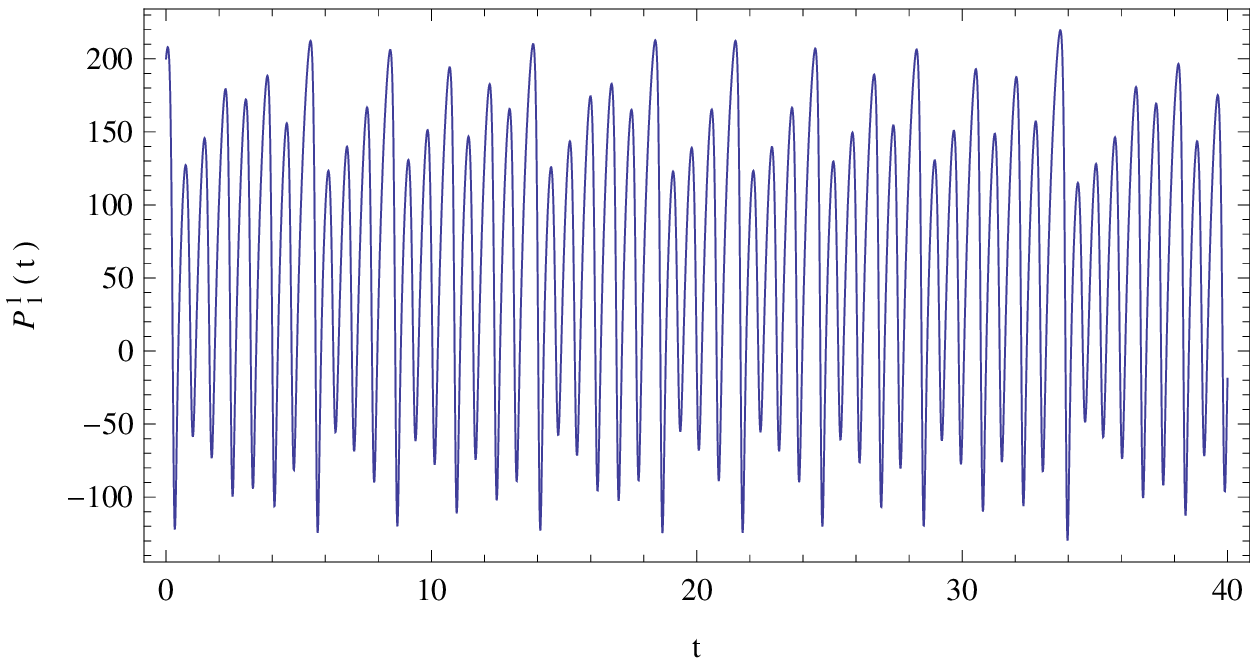}
\includegraphics[width=8cm]{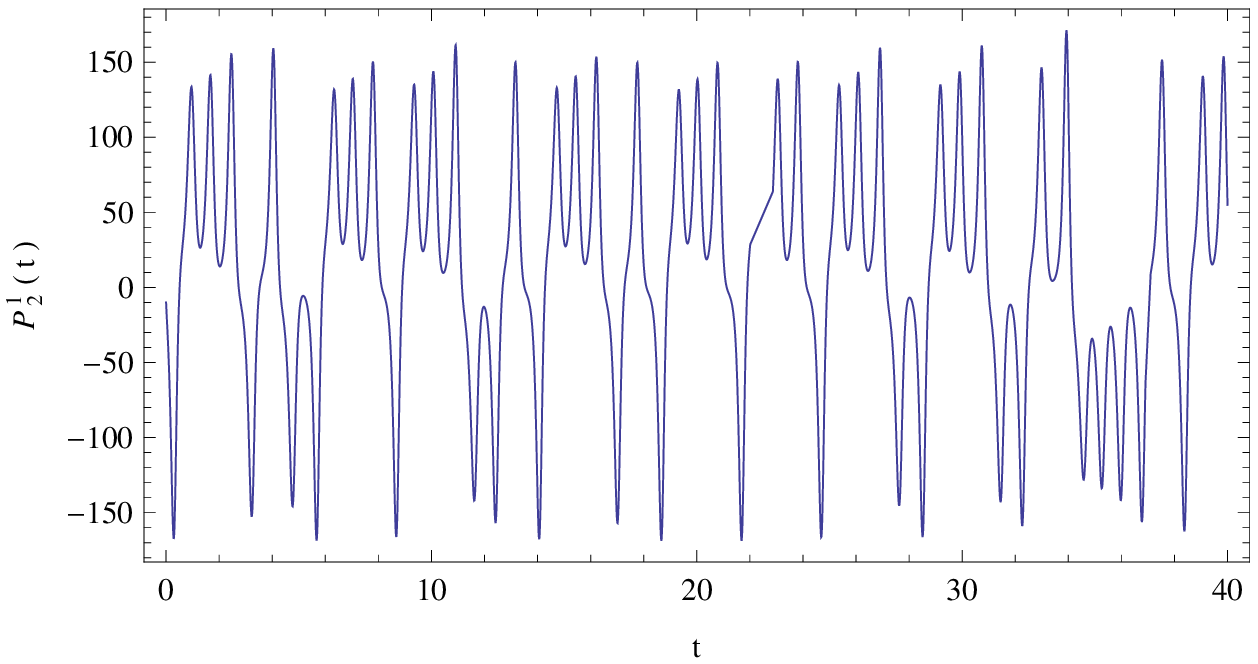}
\end{centering}
\caption{Time variation of the deviation curvature tensor component $P_1^1(t)$, shown in the left figure, and the
time variation of $P_2^1(t)$, presented in the right figure, for $\sigma =10$, $\rho =28$, and $\beta =8/3$, respectively. The initial conditions used for the numerical integration of the Lorenz system are $X(0)=1$, $Y(0)=5$, and $Z(0)=10$. }\label{Lor3}
\end{figure*}

\begin{figure*}[htp]
\begin{centering}
\includegraphics[width=8cm]{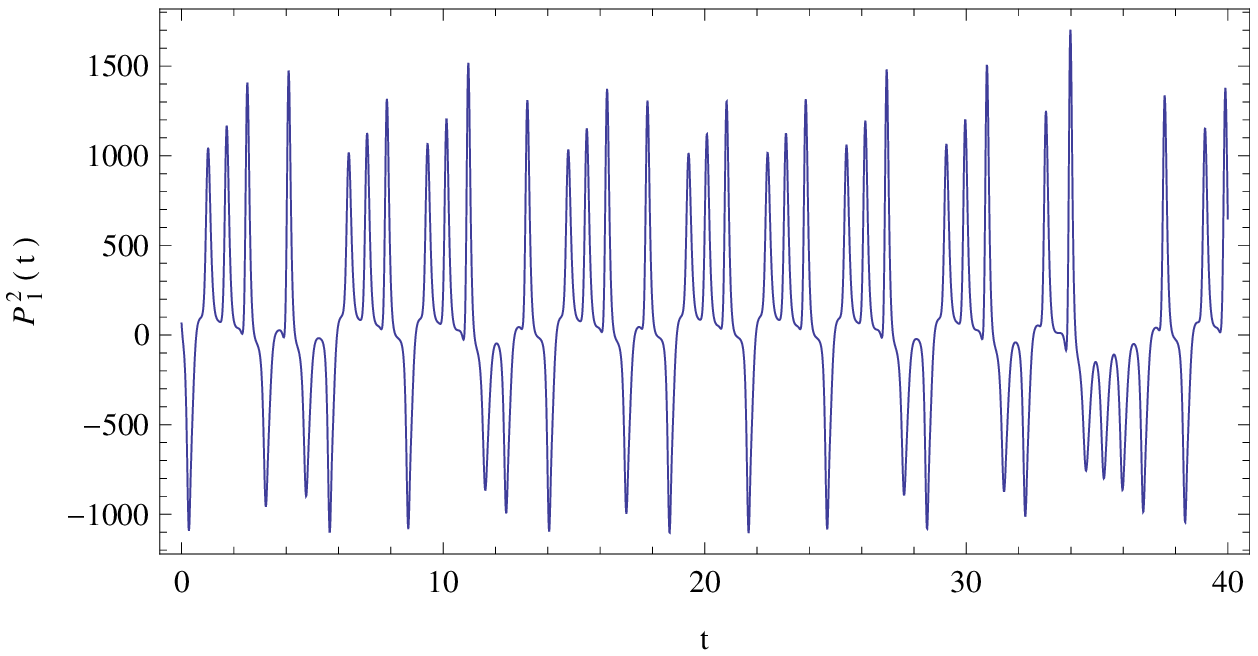}
\includegraphics[width=8cm]{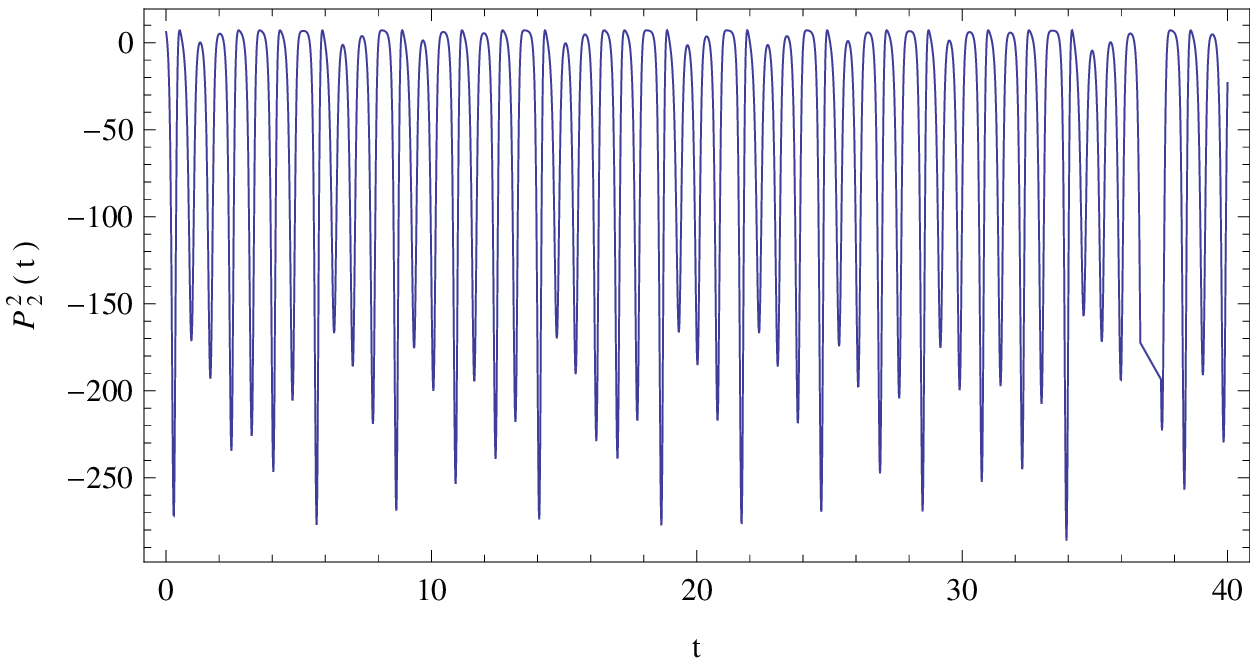}
\end{centering}
\caption{Time variation of the deviation curvature tensor component $P_1^2(t)$, presented in the left figure, and the
time variation of $P_2^2(t)$, shown in the right figure, for $\sigma =10$, $\rho =28$, and $\beta =8/3$, respectively. The initial conditions used for the numerical integration of the Lorenz system are $X(0)=1$, $Y(0)=5$, and $Z(0)=10$.}\label{Lor4}
\end{figure*}

The time variation of the components of the deviation curvature tensor for the Lorenz system are represented in Figs.~\ref{Lor3}-\ref{Lor4}.

The third, fourth and fifth KCC invariants, as defined by Eqs.~(\ref{31}) are identically equal to zero for the Lorenz system. However, the third invariant as defined by Eq.~(\ref{32}) has a non-zero component. Generally, the third KCC invariant can be written as
\begin{equation}
B_{jk}^{i}=\frac{\partial N_{j}^{i}}{\partial X^{k}}-\frac{\partial N_{k}^{i}%
}{\partial X^{j}}+N_{j}^{m}\frac{\partial N_{k}^{i}}{\partial Y^{m}}%
-N_{k}^{l}\frac{\partial N_{j}^{i}}{\partial Y^{l}},
\end{equation}
which has a single non-zero component,
\be
B_{12}^2=-N^1_2\frac{\partial N_1^2}{\partial Y^1}=-\frac{1+\sigma }{2\sigma }.
\ee

All the other KCC invariants of the Lorenz system are identically equal to zero.

\subsection{The Jacobi stability of the equilibrium points of the Lorenz system}

The three equilibrium points of the Lorenz system are $S_0(0,0,0)$, if $\rho \leq 1$,   \\
$S_{+}\left[\sqrt{\beta \left(\rho -1\right)},\sqrt{\beta \left(\rho -1\right)},\rho -1\right]$, and $S_{-}\left[-\sqrt{\beta \left(\rho -1\right)},-\sqrt{\beta \left(\rho -1\right)},\rho -1\right]$, if $\rho >1$, respectively. From the point of view of the second order differential formulation of the Lorenz system and of the Jacobi analysis, the equilibrium points of the system given by Eqs.~(\ref{Ff1}) and (\ref{Ff2}) are
\be
S_0\left(X^1_{0}=0, X^2_0=0\right), \rho \leq 1,
\ee
\be
S_{+}\left[X^1_{+}=\sqrt{\beta (\rho -1)},X_{+}^2=\rho -1\right], \rho >1,
\ee
and
\be
S_{-}\left[X^1_{-}=-\sqrt{\beta (\rho -1)},X_{-}^2=\rho -1\right], \rho >1,
\ee
respectively.

In the equilibrium points the components of the first KCC invariant vanish identically, so that
\be
\epsilon ^i\left(S_0\right)=\epsilon ^i\left(S_{+}\right)=\epsilon ^i\left(S_{-}\right)\equiv 0, i=1,2.
\ee

We evaluate now the components of the curvature deviation tensor in the equilibrium points. First we obtain
\be
P_1^1\left(S_0\right)=-(1-\rho)\sigma +\frac{1}{4}\left(1+\sigma\right)^2,  \rho \leq 1
\ee
\be
P_1^2\left(S_0\right)=P_2^1\left(S_0\right)=0,  \rho \leq 1
\ee
\be
P_2^2\left(S_0\right)=\beta ^2, \rho \leq 1,
\ee

For the equilibrium points $S_{+}$ and $S_{-}$ we obtain
\be
P_1^1\left(S_{+}\right)=\frac{1}{4} (1+\sigma )^2, \rho >1,
\ee
\be
P_2^1\left(S_{+}\right)=  -\sigma\sqrt{\beta  (\rho -1)},\rho >1,
\ee
\be
P_1^2\left(S_{+}\right)=\frac{\sqrt{\beta  (\rho -1)} \left(-7 \beta  \sigma +\beta -\sigma ^2+1\right)}{4 \sigma }, \rho >1,
\ee
\be
P_2^2\left(S_{+}\right)=\beta ^2-\beta  \left(\rho -1\right), \rho >1,
\ee
and
\be
P_1^1\left(S_{-}\right)=\frac{1}{4} (1+\sigma )^2, \rho >1,
 \ee
 \be
 P^1_2\left(S_{-}\right)=\sigma\sqrt{\beta  (\rho -1)},\rho >1,
 \ee
 \be
P^2_1\left(S_{-}\right)=\frac{\sqrt{\beta  (\rho -1)} \left(\beta  (7 \sigma -1)+\sigma ^2-1\right)}{4 \sigma }, \rho >1,
\ee
\be
P_2^2\left(S_{-}\right)=\beta ^2-\beta  \left(\rho -1\right), \rho >1,
\ee
respectively.

The eigenvalues of the curvature deviation tensor in the equilibrium
points are obtained as \be
\label{lambaSzero}
\lambda
_{+}\left(S_0\right)=\frac{1}{4} \left[\sigma  (4 \rho +\sigma
-2)+1\right], \lambda _{-}\left(S_0\right)=\beta ^2, \rho \leq 1,
\ee
\begin{widetext}
\bea
\lambda _{+}\left(S_{+}\right)&=&\frac{1}{2} \Bigg\{\sqrt{\left[\beta \left(\beta-  \rho +1\right) -\frac{1}{4} (\sigma +1)^2\right]^2+\beta  (\rho -1) \left[\beta  (7 \sigma -1)+\sigma ^2-1\right]}+\nonumber\\
&&\beta  (\beta
   -\rho +1)+\frac{1}{4} (\sigma +1)^2\Bigg\}, \rho >1,
\eea
\bea
\lambda _{-}\left(S_{+}\right)&=&\frac{1}{2} \Bigg\{- \sqrt{\left[\beta \left(\beta -  \rho +1\right) -\frac{1}{4} (\sigma +1)^2\right]^2+\beta  (\rho -1) \left[\beta  (7 \sigma -1)+\sigma ^2-1\right]}+ \nonumber\\
&&\beta \left(\beta
  -\rho +1\right))+\frac{(\sigma +1)^2}{4}\Bigg\}, \rho >1,
\eea
\bea
\lambda _{+}\left(S_{-}\right)&=&\frac{1}{2} \Bigg\{\sqrt{\left[\beta \left( \beta-  \rho +1\right) -\frac{1}{4} (\sigma +1)^2\right]^2+\beta  (\rho -1) \left(\beta  (7 \sigma -1)+\sigma ^2-1\right)}+\nonumber\\
&&\beta  (\beta
   -\rho +1)+\frac{1}{4} (\sigma +1)^2\Bigg\}, \rho >1,
   \eea
   \bea
  \lambda _{-}\left(S_{-}\right) &=&\frac{1}{2} \Bigg\{-\sqrt{\left[\beta \left(\beta -  \rho +1\right) -\frac{1}{4} (\sigma +1)^2\right]^2+\beta  (\rho -1) \left(\beta  (7 \sigma -1)+\sigma ^2-1\right)}+\nonumber\\
  &&\beta  (\beta
   -\rho +1)+\frac{1}{4} (\sigma +1)^2\Bigg\}, \rho >1.
   \eea
   \end{widetext}

   The eigenvalues of the deviation curvature tensor have the property
   \be
   \lambda _{+}\left(S_{+}\right)=\lambda _{+}\left(S_{-}\right),  \lambda _{-}\left(S_{+}\right)=\lambda _{-}\left(S_{-}\right).
   \ee

   The trace $\kappa $ of the deviation curvature tensor, as well as the anisotropy $\theta $ of the Lorenz system are obtained as
   \be
   \kappa \left(S_0\right)=\frac{1}{2} \left\{\beta ^2+\frac{1}{4} \left[\sigma  (4 \rho +\sigma -2)+1\right]\right\},
   \ee
   \be
    \theta \left(S_0\right)=\frac{1}{2} \left\{-\beta ^2+\frac{1}{4} [\sigma  (4 \rho +\sigma -2)+1]\right\},
   \ee
   \be
   \kappa \left(S_{+}\right)=\kappa \left(S_{-}\right)=\frac{1}{8} \left[4 \beta  (\beta -\rho +1)+(\sigma +1)^2\right],
   \ee
   and
   \begin{widetext}
   \be
   \theta \left(S_{+}\right)=\theta \left(S_{-}\right)=\frac{1}{2} \sqrt{\left[\beta \left(\beta -  \rho +1\right) -\frac{1}{4} (\sigma +1)^2\right]^2+\beta  (\rho -1) \left[\beta  (7 \sigma -1)+\sigma ^2-1\right]},
   \ee
   \end{widetext}
  respectively.

  Taking into account the previous results we can formulate the following theorem, giving the Jacobi  properties of the equilibrium points of the Lorenz system:

  {\bf Theorem.} a) The equilibrium point $S_0\left(0, 0\right)$ of the Lorenz system is Jacobi unstable.

  b) If the free parameters $\beta $, $\rho >1$, and $\sigma $ of the Lorenz system satisfy simultaneously the constraints
  \be
  \beta \left(\beta -  \rho +1\right) +\frac{1}{4} (\sigma +1)^2<0,
  \ee
  and
  \be
  \frac{1}{4} \beta  \left\{\beta  \left[-7 \rho  \sigma +\rho +\sigma  (\sigma +9)\right]-2 \sigma (\rho -1)  (\sigma +1)\right\}>0,
  \ee
  respectively, then the equilibrium points $S_{+}\left[\sqrt{\beta (\rho -1)},\rho -1\right]$ and $S_{+}\left[-\sqrt{\beta (\rho -1)},\rho -1\right]$ of the Lorenz system are Jacobi stable, and Jacobi unstable otherwise.




 \section{The onset of chaos in the Lorenz system}\label{sect5}

  The behavior of the deviation vector $\xi ^i$, $i=1,2$, giving the behavior of the trajectories of a dynamical system near a fixed point $x^i\left(t_0\right)$ is described Eqs.~(\ref{def}) and (\ref{JE}). In the case of the Lorenz system these equations can be written generally as
\bea\label{dev1}
\frac{d^2\xi ^1(t)}{dt^2}+(\sigma +1)\frac{d\xi ^1(t)}{dt}+ \sigma \left[(1-\rho ) + X_2\right]\xi^1(t)+
&\sigma  X_1 \xi ^2(t)=0,
\eea
and
\bea\label{dev2}
&&\sigma \frac{d^2 \xi ^2(t)}{dt^2}+ \left[\left(\beta  -\sigma  +1\right)X_1-2 Y_1\right]\frac{d\xi ^1(t)}{dt}+
\left[2 \sigma   (\beta -\rho +X_2+1)X_1+ (\beta
   -\sigma +1)Y_1\right]\xi ^1(t)+\nonumber\\
 &&  \sigma   (X_1-\beta ) (\beta +X_1)\xi ^2(t)=0,
\eea
respectively. The deviation vector is obtained from its components as
\be
\xi (t)=\sqrt{\left[\xi ^1(t)\right]^2+\left[\xi ^2(t)\right]^2}.
\ee

In order to obtain a quantitative description of the onset of chaos in the Lorenz system, we introduce, in analogy with the Lyapounov exponent, the instability exponents $\delta _i$, $i=1,2$, and $\delta $, defined as
\be
\delta _i(S)=\lim _{t\rightarrow \infty}\frac{1}{t}\ln\left[\frac{\xi ^i (t)}{\xi _{i0}}\right],i=1,2.
\ee
and
\be
\delta (S)=\lim _{t\rightarrow \infty}\frac{1}{t}\ln\left[\frac{\xi  (t)}{\xi _{10}}\right].
\ee

In the following we investigate the behavior of the solutions of Eqs.~(\ref{dev1}) and (\ref{dev2}) near the critical points of the Lorenz system.

\subsection{Behavior of the deviation vector near $S_0(0,0)$}

Near the equilibrium point $S_0(0,0)$ the deviation equations Eqs.~(\ref{dev1}) and (\ref{dev2}) take the form
\be
\frac{d^2\xi ^1(t)}{dt^2}+(\sigma +1) \frac{d\xi ^1(t)}{dt}+(1-\rho ) \sigma  \xi ^1(t)=0,
\ee
and
\be
  \frac{d^2\xi ^2(t)}{dt^2}-\beta ^2   \xi ^2(t)=0,
\ee
respectively. In this case the deviation equation for the origin of the Lorenz system can be separated in two independent equations, with the general solutions given by
\be\label{sol00}
\frac{\xi ^1(t)}{\xi _{10}}=\frac{e^{\frac{1}{2}  \left(\sqrt{4 \rho  \sigma +\left(\sigma -1\right)^2
   }-\sigma -1\right)t} -e^{\frac{1}{2}  \left(-\sqrt{4 \rho  \sigma +(\sigma -1)^2 }-\sigma -1\right)t}}{\sqrt{4 \rho  \sigma +(\sigma -1)^2 }},
   \ee
   and
   \be\label{sol01}
  \xi^2(t)=\frac{ e^{-\beta  t} \left(e^{2 \beta  t}-1\right)}{2 \beta }\xi _{20},
   \ee
   where we have used the initial conditions $\xi^1(0)=0,\dot{\xi}^1(0)=\xi_{10}$, and  $\xi^2(0)=0,\dot{\xi}^2(0)=\xi_{20}$, respectively. The time behavior of $\xi ^2 (t)$ is determined only by the coefficient $\beta $ of the Lorenz system. For the deviation vector we obtain
   \bea
   \xi (t)&=&\Bigg[\frac{\xi _{20}^2}{\xi _{10}^2}\frac{ \sinh ^2(\beta  t)}{4\beta ^2}+
   \frac{ e^{- \left(\sqrt{4\rho \sigma +(\sigma -1)^2}+\sigma +1\right)t} \left(e^{ \sqrt{4\rho \sigma +(\sigma -1)^2}t}-1\right)^2}{4\rho \sigma +(\sigma -1)^2}\Bigg]^{1/2}.
   \eea

\begin{figure*}[htp]
 \begin{centering}
\includegraphics[width=8cm]{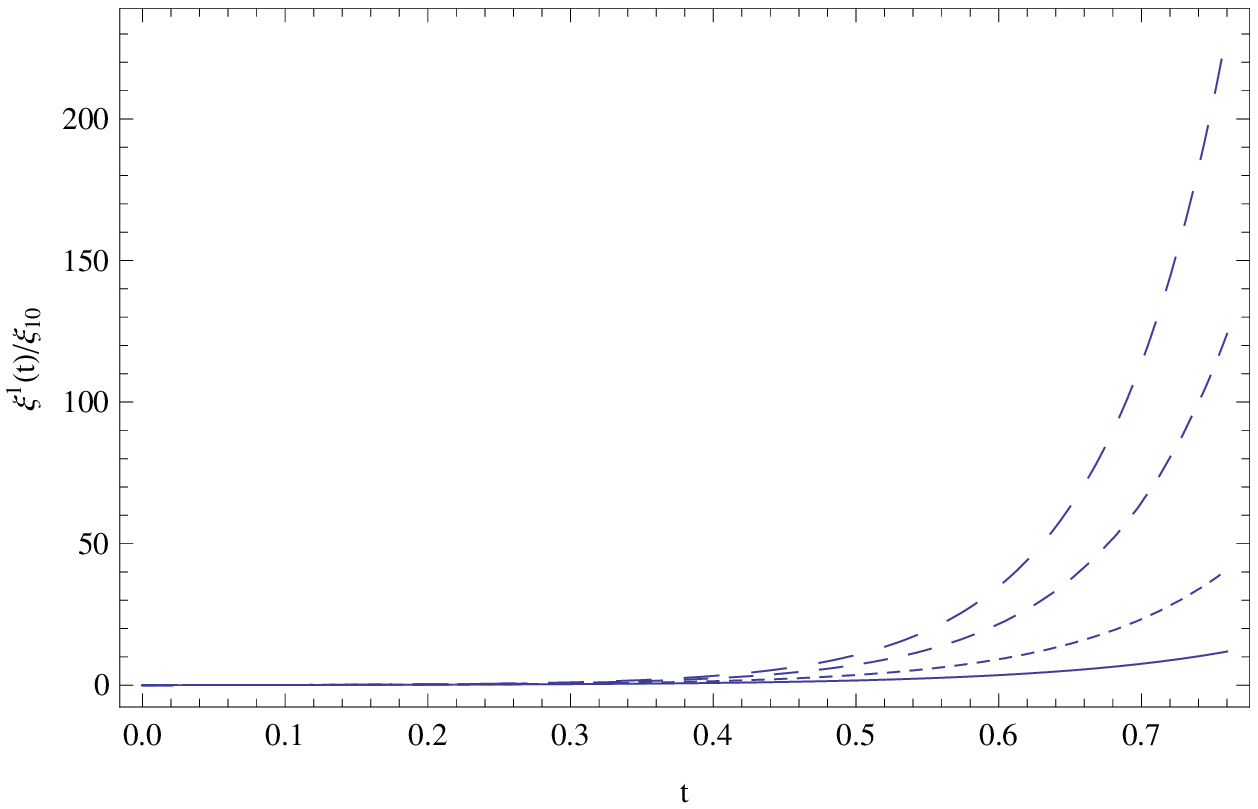}
\includegraphics[width=8cm]{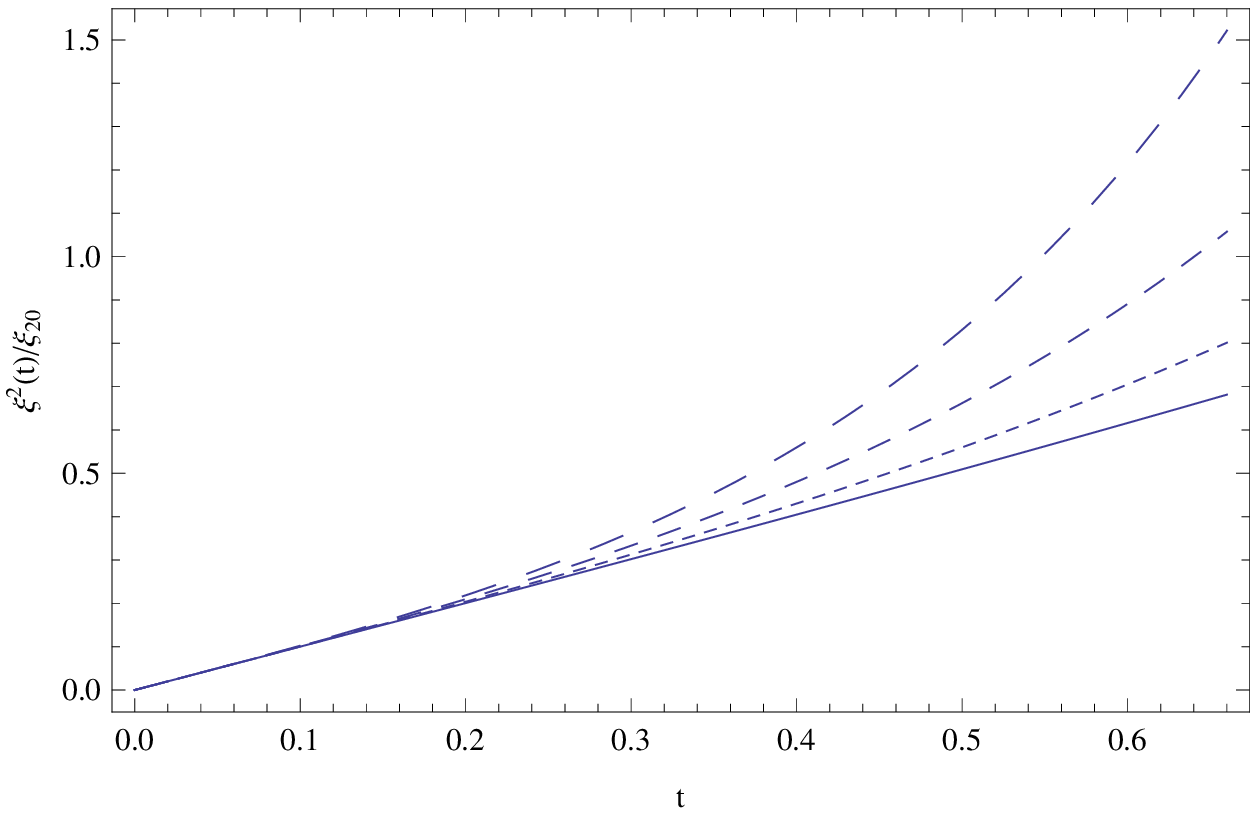}\\
\end{centering}
\caption{Time variation of the deviation vector components $\xi ^1(t)/\xi _{10}$ (left figure) and $\xi ^2 (t)/\xi_{20}$ (right figure), in the vicinity of the equilibrium point $S_0$,  for different values of the parameters $\beta $, $\rho $ and $\sigma $. In the left figure $\sigma =10$, $\beta =8/3$, and $\rho =15$ (solid curve), $\rho =20$ (dotted curve), $\rho =25$ (dashed curve), and $\rho =28 $ (long dashed curve), respectively. In the right figure $\beta =2/3$ (solid curve), $\beta =5/3$ (dotted curve), $\beta =8/3$ (dashed curve), and $\beta =11/3$ (long dashed curve), respectively. }\label{csi0}
\end{figure*}

   The time dependence of the deviation vectors $\xi ^1$ and $\xi ^2$ is represented, for different values of the parameters $\sigma $, $\rho $, and $\beta $,  in Fig.~\ref{csi0}.

\begin{figure}[htp]
 \begin{centering}
\includegraphics[scale=0.68]{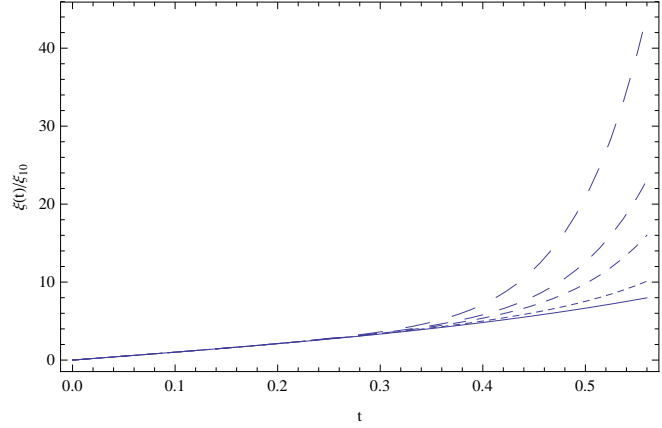}
\end{centering}
\caption{Time variation of the absolute value of the deviation vector  $\xi (t)/\xi _{10}$  for  $\beta =8/3$, $\sigma =10$, $\xi _{10}=10^{-9}$, $\xi _{20}=10^{-8}$, and for different values of $\rho $: $\rho =10$ (solid curve), $\rho =20$ (dotted curve), $\rho =25$ (dashed curve), $\rho =28 $ (long dashed curve), and $\rho =33$ (ultra-long dashed curve, respectively. }\label{csitot0}
\end{figure}

For a fixed $\beta $ and $\sigma $, with the increase of the parameter $\rho $, the deviation curvature component $\xi ^1$ increases very rapidly in time, indicating the onset of chaos in the Lorenz system. In the large time limit the increase of $\xi ^1(t)$ is exponential. The deviation curvature vector component $\xi ^2$ also increases exponentially in time, a behavior that is independent of the values of $\beta $. The time variation of the absolute value of the deviation vector $\xi (t)$ is represented in Fig.~\ref{csitot0}.

With the increase of $\rho $ the absolute value of the deviation vector rapidly increases in time, indicating the onset of chaos in the Lorenz system.
With the help of Eqs.~(\ref{sol00}) and (\ref{sol01}) we immediately obtain
\be
\delta _1\left(S_0\right)=\frac{1}{2}\left(\sqrt{4\rho \sigma +\left(\sigma -1\right)^2}-\sigma -1\right),
\ee
and
\be
\delta _2\left(S_0\right)=\beta,
\ee
respectively. The instability exponent $\delta $ can be estimated as
\be
\delta =\frac{1}{2t}\ln \left( \frac{\xi _{20}^{2}}{\xi _{10}^{2}}\frac{%
e^{2\beta t}}{4\beta ^{2}}+\frac{e^{\sqrt{4\rho \sigma +(\sigma -1)^{2}}t}}{%
4\rho \sigma +(\sigma -1)^{2}}\right) .
\ee

\begin{figure}[htp]
 \begin{centering}
\includegraphics[scale=0.68]{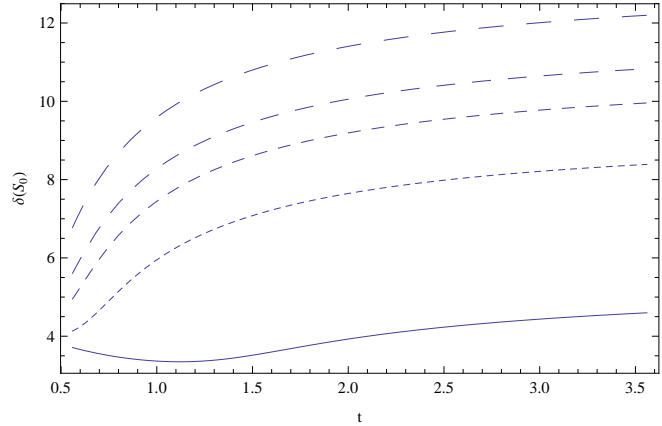}
\end{centering}
\caption{Time variation of the instability exponent $\delta $  for  $\beta =8/3$, $\sigma =10$, $\xi _{10}=10^{-9}$, $\xi _{20}=10^{-8}$, and for different values of $\rho $: $\rho =10$ (solid curve), $\rho =20$ (dotted curve), $\rho =25$ (dashed curve), $\rho =28 $ (long dashed curve), and $\rho =33$ (ultra-long dashed curve, respectively. }\label{deltaS0}
\end{figure}

The time variation of the instability exponent $\delta $ is represented, for fixed values of the parameters $\beta $ and $\sigma $, in Fig.~\ref{deltaS0}.

\subsection{Dynamics of the deviation vector near $S_{+}\left[\sqrt{\beta (\rho -1)},\rho -1\right]$ and $S_{-}\left[-\sqrt{\beta (\rho -1)},\rho -1\right]$}

   For both fixed points  $S_{+}\left[\sqrt{\beta (\rho -1)},\rho -1\right]$ and $S_{-}\left[-\sqrt{\beta (\rho -1)},\rho -1\right]$,  the differential equations describing the dynamics of the deviation vector near the given fixed points take the form
   \begin{widetext}
   \be
   \frac{d^2\xi ^1(t)}{dt^2}+(\sigma +1) \frac{d\xi ^1(t)}{dt}+\sigma\sqrt{\beta } \sqrt{\rho -1}   \xi ^2(t)=0,
   \ee
   and
   \be
   \frac{d^2\xi ^2(t)}{dt^2}+\frac{\sqrt{\beta } \sqrt{\rho -1} (\beta -\sigma +1) }{\sigma }\frac{d\xi ^1(t)}{dt}+2 \beta ^{3/2} \sqrt{\rho -1} \xi ^1(t)-\beta  (\beta -\rho +1) \xi   ^2(t)=0,
   \ee
   \end{widetext}
respectively.

\begin{figure*}[htp]
\centering
\includegraphics[width=8cm]{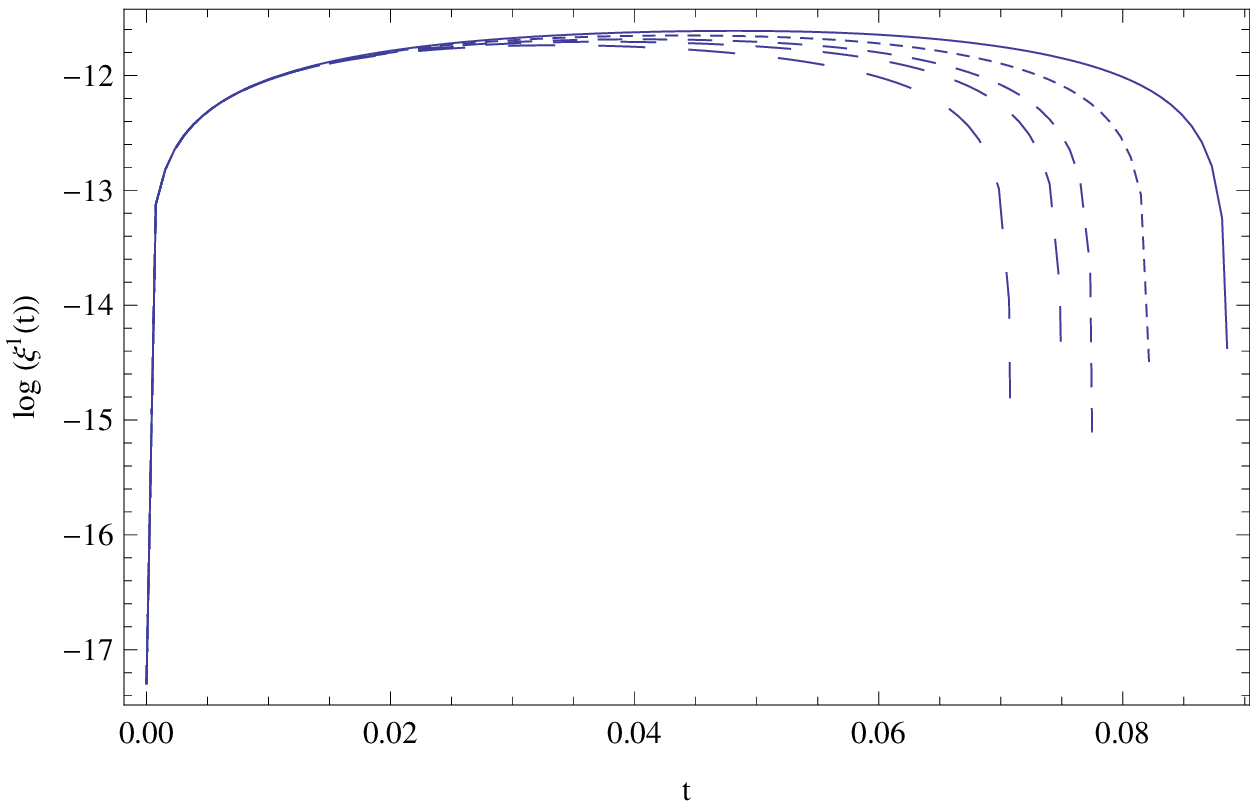}
\includegraphics[width=8cm]{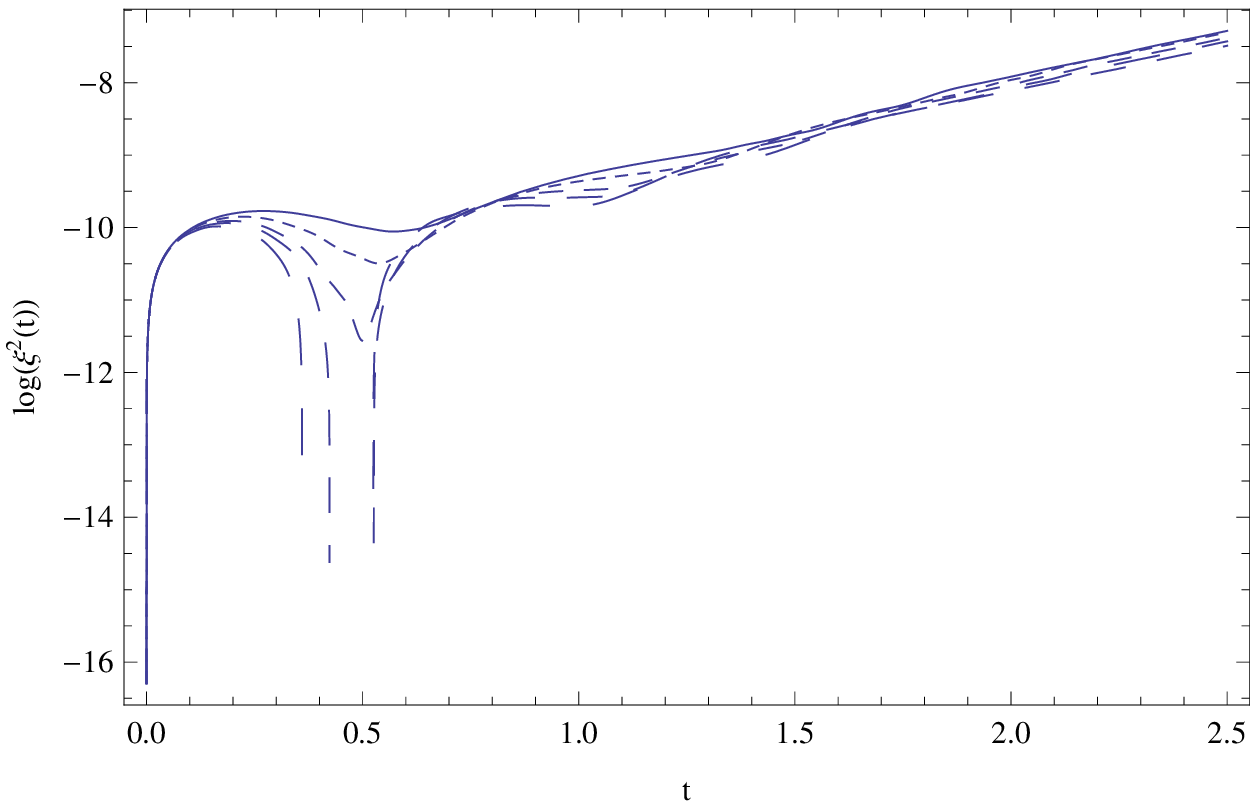}\\
\caption{Time variation of the deviation vector components $\xi ^1(t)$ (left figure) and $\xi ^2 (t)$ (right figure) in a logarithmic scale near the equilibrium points $S_{\pm}\left(\pm \sqrt{\beta (\rho -1)}\right)$, for  $\beta =8/3$,  $\sigma =10$, and for different values of $\rho $: $\rho =15$ (solid curve), $\rho =20$ (dotted curve), $\rho =25$ (dashed curve),  $\rho =28 $ (long dashed curve), and $\rho =33 $ (ultra-long dashed curve, respectively. The initial conditions used to integrate the deviation equations are $\xi ^1 (0)=\xi ^2(0)=0$, $\dot{\xi}^1(0)=10^{-10}$, and $\dot{\xi }^2 (0)=10^{-9}$, respectively.  }\label{csip1}
\end{figure*}

The behavior of the components of the deviation curvature vector is shown in Fig.~\ref{csip1}.

\begin{figure}[htp]
\centering
\includegraphics[scale=0.68]{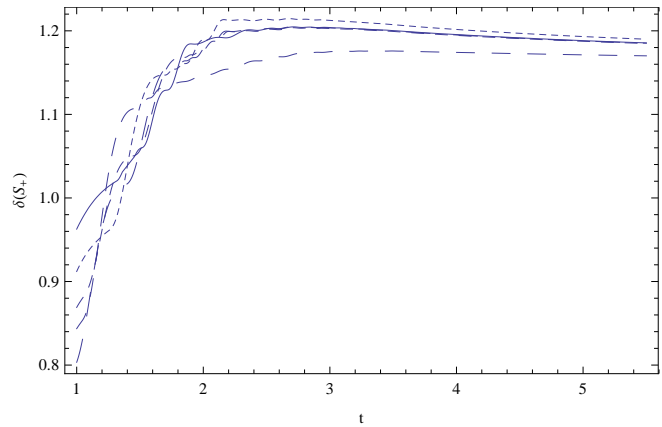}
\caption{Time variation of the instability exponent $\delta \left(S_{+}\right)=\delta \left(S_{-}\right)$ near the equilibrium points $S_{\pm}\left(\pm \sqrt{\beta (\rho -1)}\right)$, for  $\beta =8/3$,  $\sigma =10$, and for different values of $\rho $: $\rho =15$ (solid curve), $\rho =20$ (dotted curve), $\rho =25$ (dashed curve),  $\rho =28 $ (long dashed curve), and $\rho =33 $ (ultra-long dashed curve, respectively. The initial conditions used to integrate the deviation equations are $\xi ^1 (0)=\xi ^2(0)=0$, $\dot{\xi}^1(0)=10^{-10}$, and $\dot{\xi }^2 (0)=10^{-9}$, respectively.  }\label{deltap}
\end{figure}

The time variation of the instability exponent $\delta \left(S_{+}\right)=\delta \left(S_{-}\right)$ is represented in Fig.~\ref{deltap}.

\subsection{The curvature of the deviation vector}

In order to obtain a quantitative description of the behavior of the curvature deviation tensor in the following we analyze the signed geometric curvature $\kappa _0$ of the curve $\xi (t)=\left(\xi ^1 (t),\xi ^2(t)\right)$, which we define, according to the standard approach used in the differential geometry of plane curves as
\be
\kappa _0(S)=\kappa _{0}=\frac{\dot{\xi}^{1}(t)\ddot{\xi}^{2}(t)-\ddot{\xi}^{1}(t)\dot{\xi%
}^{2}(t)}{\left\{ \left[ \dot{\xi}^{1}(t\right] ^{2}+\left[ \dot{\xi}^{2}(t%
\right] ^{2}\right\} ^{3/2}}.
\ee

\begin{figure}[h]
\centering
\includegraphics[scale=0.68]{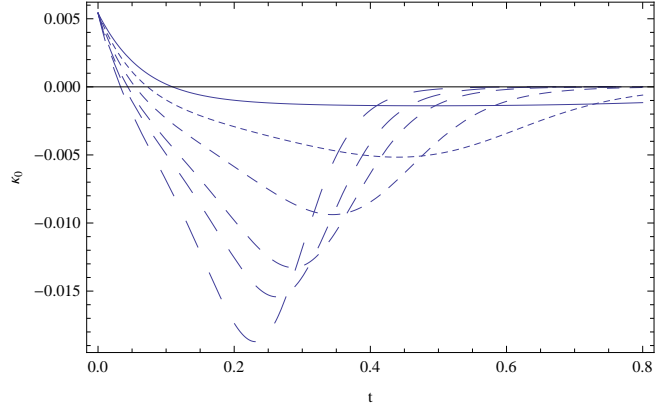}
\caption{Time variation of the curvature $\kappa _0$ of the deviation vector $\xi (t)$  near the equilibrium points $S_{0}\left(0,0\right)$, for  $\beta =8/3$,  $\sigma =10$, $\dot{\xi}^1(0)=10^{-10}$, and $\dot{\xi }^2 (0)=10^{-9}$, respectively, and for different values of $\rho $: $\rho =10$ (solid curve), $\rho =15$ (dotted curve), $\rho =20$ (dashed curve),  $\rho =25 $ (long dashed curve),  $\rho =28 $ (ultra-long dashed curve, and $\rho =33$ (ultra-ultra-long dashed curve), respectively. }\label{curv}
\end{figure}

If we denote $\sqrt{4\rho\si+(\si-1)^2}=a$ and $\si+1=b$, we obtain an
explicit formula for $\kappa_0$,
\begin{widetext}
\begin{eqnarray}
\kappa _{0}&=&\frac{\xi _{10}\xi _{20}}{8a}\frac{e^{(-b/2)t}}{\left\{ \frac{%
\xi _{10}^{2}}{4a^{2}}[(a-b)e^{(a/2)t}+(a+b)e^{(-a/2)t}]^{2}e^{-bt}+\frac{%
\xi _{20}^{2}}{4}[e^{\beta t}+e^{-\beta t}]^{2}\right\} ^{3/2}}\times \nonumber\\
&&\Bigg[
(a-b)(2\beta -a+b)e^{(a/2+\beta )t}+(a-b)(b-2\beta -a)e^{(a/2-\beta
)t}+\nonumber\\
&&(a+b)(2\beta +a+b)e^{(-a/2+\beta )t}+(a+b)(a+b-2\beta )e^{(-a/2-\beta
)t}\Bigg] .
\end{eqnarray}
\end{widetext}
In the following we restrict our study to the equilibrium point
$S_0(0,0)$, and we fix the values of the parameters as $\sigma =10$,
$\beta =8/3$, $\xi _{10}=10^{-10}$, and $\xi _{20}=10^{-9}$,
respectively. The variation of the curvature $\kappa _0$ is
represented in Fig.~\ref{curv}.

As one can see from the figure, for the chosen range of the parameters the curvature of the two-dimensional deviation vector is positive for small values of time, reaches the value zero at a certain moment $t_0$, enters the region of the negative values, and then it tends to zero in the limit of large times. Near the time origin $t=0$ the curvature of the deviation vector curve can be represented in a form of a power series as
\bea
\kappa _0(t,\rho)&\approx \frac{11}{202 \sqrt{101}}-\frac{4545 \sqrt{101} \rho +45584 \sqrt{101} }{9272709}t+
\frac{11 \left(877185 \sqrt{101} \rho +2748853 \sqrt{101}\right)
  }{1873087218} t^2+...
   \eea
In the first approximation the time interval $t_0$ for which $\kappa _0\left(t_0,\rho \right)=0$ is given by
\be
t_0\approx \frac{9999}{2 (4545 \rho +45584)}\approx \frac{1.099}{\rho +10.02}.
\ee

Therefore the time interval after which the curvature $\kappa _0$ changes sign is an indicator of the development of chaos in the Lorenz system. There exist therefore a critical time interval $t_0^{crit}$ so that if the transition from positive to negative $\kappa _0$ occurs at times $t>t_0^{crit}$, the evolution of the Lorenz system is predictable, and deterministic, while if the transition occurs for time intervals $t<t_0^{crit}$, then the underlying evolution of the Lorenz system will be chaotic in the long term.  The curvature becomes zero when the two components of the deviation vector satisfy the condition $\xi ^1\left(t_0\right)=\xi ^2 \left(t_0\right)$.

\subsection{Comparing with linear stability analysis}

We will now compare Jacobi stability and linear stability at the
equilibrium point $S_0(0,0)$. The linearized Lorenz system at
$(0,0,0)$ is given by

$$
\left[
  \begin{array}{c}
    \dot{X} \\
    \dot{Y} \\
    \dot{Z} \\
  \end{array}
\right] =\left[
   \begin{array}{ccc}
     -\si & \si & 0 \\
     \rho& -1 & 0 \\
     0 & 0 & -\bee \\
   \end{array}
 \right]
 \left[
  \begin{array}{c}
    X \\
    Y \\
    Z \\
  \end{array}
\right].
$$

We see that the equation for $Z$ is decoupled and $Z(t)$ tends to
$0$ exponentially fast. The time dynamics of $X$ and $Y$ is governed by the two-dimensional system

$$
\left[
  \begin{array}{c}
    \dot{X} \\
    \dot{Y} \\
  \end{array}
\right] =\left[
   \begin{array}{cc}
     -\si & \si \\
     \rho& -1 \\
   \end{array}
 \right]
 \left[
  \begin{array}{c}
    X \\
    Y \\
  \end{array}
\right].
$$
Let $A:=\left[
   \begin{array}{cc}
     -\si & \si \\
     \rho& -1 \\
   \end{array}
 \right]$. Let $\tau=\mbox{trace}(A)$ and $\Delta=\det(A)$.
Then,
$$
\lambda_1=\frac{\tau+\sqrt{\tau^2-4\Delta}}{2},\quad
\lambda_2=\frac{\tau-\sqrt{\tau^2-4\Delta}}{2}.
$$
On the other  hand, in Jacobi stability analysis, we find the
eigenvalues of $P^i_j(0,0)$ are \bea \lambda _{+}\left(S_0\right)&=&
\frac{1}{4} \left[\sigma  (4 \rho +\sigma
-2)+1\right]=\frac{1}{4}(\tau^2-4\Delta),\\
\lambda _{-}\left(S_0\right)&=& \beta^2, \rho \leq 1, \eea

Thus, one of eigenvalues of $P^i_j(0,0)$ recovers some information
of the stability of the system at the origin. That is, the origin
is a center or spiral, when $\lambda
_{+}\left(S_0\right)<0$.

\section{Discussions and final remarks}\label{sect6}

In the present paper we have considered the stability analysis of the Lorenz system from the point of view of the KCC theory, in which the dynamical stability properties of dynamical systems are inferred from the study of the geometric properties of the Finsler space geodesic equations, equivalent with the given system. By transforming the Lorenz to an equivalent system of two second order differential equations, the dynamical system can be interpreted as representing the geodesic motion of a "particle" in an associated Finsler space. This geometrization of the Lorenz system opens the possibility of applying the standard methods of differential geometry for the study of its properties. We have obtained, and analyzed in detail, the main geometrical objects that can be associated to the Lorenz system, namely, the non-linear connection, the Berwald connection, and the first, second and third KCC invariants. The main result of the present paper is the Jacobi stability condition of the equilibrium points of the Lorenz system, showing that the origin is always Jacobi unstable, while  the Jacobi stability of the other two equilibrium points depends on the values of the parameters of the system. By considering the standard values of $\sigma $, $\beta $ and $\rho $ in the Lorenz system, $\sigma =10$, $\beta =8/3$, and $\rho =28$, it turns out that for this choice of the parameters all the equilibrium points are Jacobi unstable. From the point of view of linear stability analysis, for $0<\rho < 1$, the zero fixed point $S_0=(0, 0, )$ is globally stable; from a physical point of view this refers to
the non convective state. For $\rho >l$ and $\rho $ not too large, the state with roll convection, referring to the $S_{\pm}$ equilibrium points, is stable \cite{And}. From a physical point of view this means that the phase space  consists of two regions, which are separated by the stable manifold of the zero fixed point; the trajectories which start in one of the two regions are attracted by the corresponding nonzero
fixed point.

We have also considered in detail the behavior of the deviation
vector near the equilibrium points. In order to describe the
behavior of the trajectories near the equilibrium points we have
introduced the instability exponent $\delta $, as well as the
curvature $\kappa _0$ of the deviation vector trajectories. The
curvature $\kappa _0$ of the curve $\xi (t)$ can be related directly
to the chaotic behavior of the trajectories via its transition
moment from positive to negative values. An early  transition
indicates the presence of chaotic states. Therefore we suggest the
use of the curvature of the deviation vector as an indicator of the
onset of chaos in non-linear dynamical systems.  In \cite{T1} it was
suggested that the torsion tensor geometrically expresses the
chaotic behavior of dynamical systems, i.e. a trajectory of
dynamical systems with the torsion tensor is not closed. By using
the same definition as in \cite{T1} it turns out that indeed there
is a non-zero torsion tensor component $B_{12}^2=-(1+\sigma
)/2\sigma \neq 0$, $\forall \sigma \neq 0$. Therefore the existence
of chaos in the Lorenz system is intimately related to a non-zero
$\sigma $, which is needed for the chaotic behavior of the Lorenz
system \cite{Pel}.

\end{document}